\documentstyle[12pt]{article}
\newcommand{\be}{\begin{equation}}
\newcommand{\ee}{\end{equation}}

\date{May 2000}
\begin{document}
\begin{center}
{\large {\bf On the relation between polynomial deformations of
$sl(2,R)$ and quasi-exactly solvability}}\\
\vspace{0.5in}
 N. DEBERGH \footnote{Chercheur, Institut Interuniversitaire des
Sciences Nucl\'eaires, Bruxelles, email: Nathalie.Debergh@ulg.ac.be}

\vspace{0.3in}
 {\it Theoretical and Mathematical Physics,\\
Institute of Physics (B5)\\
University of Li\`ege,\\
B-4000 Li\`ege 1 (Belgium)}\\
\vspace{0.2in}
\end{center}
\vspace{0.5in}
\begin{abstract}
A general method based on the polynomial deformations of the Lie
algebra $sl(2,R)$ is proposed in order to exhibit the quasi-exactly
solvability of specific Hamiltonians implied by quantum physical
models. This method using the finite-dimensional representations and
differential realizations of such deformations is illustrated on the
sextic oscillator as well as on the second harmonic generation.
\end{abstract}
\newpage
\section{Introduction}
\hspace{5mm}
In the late eighties, Turbiner and Ushveridze [1] discovered some
cases where a finite number of eigenvalues (and the associated
eigenfunctions) of the spectral problem for the Schr\" odinger operator
\be
\bar H \psi = E \psi \; , \; \bar H = - \frac{d^2}{dy^2} + V(y) \; ,
\; y \in R \; or \; R^+
\ee
can be found explicitely. The corresponding problems have been called
quasi-exactly solvable (QES). Since that first step, QES equations
have been classified [2] according to their relation with the
finite-dimensional representations of the Lie algebra $sl2,R)$. Indeed
QES Schr\" odinger equations as given in (1) can be written as
\be
H \phi = E \phi \; , \; H = p_4(x) \frac{d^2}{dx^2} + p_3(x)
\frac{d}{dx} + p_2(x) \; ,\; x \in R \; or \; R^+
\ee
through ad-hoc changes of variables and functions [2]
\be
x = x(y) \; , \; \psi = exp(\chi) \phi,
\ee
if $p_j(x) (j=2,3,4)$ refer to polynomials of order $j$ in $x$. The
Hamiltonian $H$ in (2) can also be expressed in terms of the
first-order differential operators
$$
j_+ = -x^2 \frac{d}{dx} + 2j x ,
$$
\be
j_0 = x \frac{d}{dx} - j,
\ee
$$
j_- = \frac{d}{dx}, \; \; j=0,\frac{1}{2},1,...,
$$
satisfying the $sl(2,R)$ commutation relations i.e.
\be
[j_0 , j_{\pm}] = \pm j_{\pm},
\ee
\be
[j_+ , j_-] = 2 j_0,
\ee
the Casimir operator of this structure being $C = j_+j_- + j_0^2
-j_0$. The operator $H$ is then
\be
H = \sum_{\mu , \nu = \pm , 0 \\ ,\mu \geq \nu} c^{\mu
\nu} j_{\mu} j_{\nu} + \sum_{ \mu = \pm , 0} c^{\mu}
j_{\mu}
\ee
where the coefficients $ c^{\mu \nu}, c^{\mu}$ are arbitrary
real numbers.
\par
The crucial point in order to relate the operator (7) to QES problems
is the introduction of the nonnegative integer $2j$ in (4). Indeed,
the generators of $sl(2,R)$ as written in (4) preserve the space of
polynomials of order $2j$
\be
P(2j) = \{ 1, x, x^2, ..., x^{2j} \}
\ee
and so do the Hamiltonian (7). Searching for the eigenvalues of (1) is
thus limited to the diagonalization of (7) in the $(2j+1)$-dimensional
space $P(2j)$. It is a straightforward problem leading to the
knowledge of the eigenvalues $E_k (k=0,1,...,2j)$ as well as the
corresponding eigenfunctions $\psi_k$ (cf. (3) where $\phi_k$ belongs to
$P(2j)$) and ensuring the quasi-exactly solvability of the original
equation.
\par
We propose in this paper to take a new look at this problem through
the consideration of the so-called polynomial deformations [3] of
$sl(2,R)$ i.e. of the structures characterized by the following
commutation relations
\be
[J_0 , J_{\pm}] = \pm J_{\pm},
\ee
\be
[J_+ , J_-] = p_n(J_0),
\ee
where $p_n(J_0)$ stands for a polynomial of order $n$ in the operator
$J_0$.
\par
More precisely, in Section 2, we show how to introduce in a natural
manner the polynomial deformations (10) inside the operator (7) (n
will then be limited to 3). In Section 3, we study the
finite-dimensional representations of these polynomial deformations
while Section 4 is devoted to their finite-dimensional differential
realizations. In Section 5, we analyze two examples i.e. the sextic
oscillator (Subsection 5.1) and the second harmonic generation (SHG)
problem (Subsection 5.2). Finally, we give some conclusions in Section
6.
\section{Polynomial deformations of $sl(2,R)$ inside QES problems}
\hspace{5mm}
Instead of considering the operators (4) as expressed in (7), let us
introduce the following operators (so defined for natural reasons in
connection with their respective raising, diagonal or lowering
characteristics)
\be
J_+ \equiv c^{++} j_+^2 + c^{+0} j_+j_0 + c^+ j_+,
\ee
\be
J_0 \equiv c^{+-} j_+j_- + c^{00} j_0^2 + c^0 j_0,
\ee
\be
J_- \equiv c^{0-} j_0j_- + c^{--} j_-^2 + c^- j_-,
\ee
so that $H$ simply writes
\be
H = J_+ + J_0 + J_-.
\ee
As we will see, restoring the linearity inside (7) such that it
becomes the combination (14) will have the consequence of introducing
nonlinearity inside (6) so that we will be concerned with the algebra
(9)-(10). Indeed asking for the relations (9) to be satisfied with the
operators (11)-(13) and the relations (5)-(6) lead to two cases only
i.e. either
\be
c^{++} = c^{--} = 0, c^{+-} = c^{00}, c^0 + c^{00} = 1
\ee
or
\be
c^{++} \neq 0, c^{--} \neq 0, c^{+0} = c^+ = c^{0-} = c^- = 0, c^{+-}
= c^{00}, c^0 + c^{00} = \frac{1}{2}.
\ee
The relation (10) is then ensured with the respective polynomials
$p_{n=3} (J_0)$
\begin{eqnarray}
&&p_3(J_0) = 4 c^{+0}c^{0-} J_0^3 + 3(c^{0-}c^+ + c^{+0}c^- -
c^{+0}c^{0-})J_0^2 \nonumber \\
&&+ [2c^+c^- - c^{+0}c^- - c^{0-}c^+ + c^{+0}c^{0-}(1-2j(j+1))] J_0 \\
&&+ j(j+1) (c^{+0}c^{0-} - c^{+0}c^- - c^{0-}c^+) \nonumber
\end{eqnarray}
or
\be
p_3(J_0) = 8 c^{++}c^{--} ((2j^2+2j-1)J_0 - 8J_0^3).
\ee
Notice that in these expressions, $c^{+-} (= c^{00})$ has been put
equal to zero without loosing generality (cf. the Casimir operator of
$sl(2,R)$). Moreover, the relation (18) refers to the Higgs algebra
already intensively visited [3].
\par
We can thus consider the QES Hamiltonians as linear combinations of
operators generating the following polynomial deformation of $sl(2,R)$
\be
[J_0 , J_{\pm}] = \pm J_{\pm},
\ee
\be
[J_+ , J_-] = \alpha J_0^3 + \beta J_0^2 + \gamma J_0 + \delta, \;
\alpha, \beta, \gamma, \delta \in R
\ee
taking account of the two possibilities (17) and (18). Notice that the
Casimir operator of this deformed algebra is $$C = J_+J_- +
\frac{\alpha}{4} J_0^4 + (\frac{\beta}{3} - \frac{\alpha}{2}) J_0^2 +
(\frac{\alpha}{4} - \frac{\beta}{2} + \frac{\gamma}{2}) J_0^2 +
(\frac{\beta}{6} - \frac{\gamma}{2} + \delta) J_0$$.
\par The next step will be the determination of the finite-dimensional
representations of the algebra (19)-(20) denoted in the following by
$sl^{(3)} (2,R)$, the upper index referring to the highest power of
the diagonal operator.
\section{Finite-dimensional representations of $sl^{(3)} (2,R)$}
\hspace{5mm}
As stated in the Introduction and in relation with the possible
diagonalization of $H$, we are interested in the finite-dimensional
($= 2J+1, J=0,\frac{1}{2},1,...$) representations of $sl^{(3)} (2,R)$,
only. We thus consider kets of type $\mid J,M>$ with $M$ running from
$-J$ to $J$ and such that
\be
J_0 \mid J,M> = (\frac{M}{q} + c) \mid J,M>,
\ee
\be
J_+ \mid J,M> = f(M) \mid J,M+q>,
\ee
\be
J_- \mid J,M> = g(M) \mid J,M-q>,
\ee
where $q$ is a positive integer and $c$ a real number. The relations
(21)-(23) are such that (19) is satisfied. In order to ensure (20), we
have to impose
\begin{eqnarray}
&&f(M-q)g(M)-f(M)g(M+q) = \alpha (\frac{M}{q} + c)^3 + \beta
(\frac{M}{q} + c)^2 \nonumber \\
&&+ \gamma (\frac{M}{q} + c) + \delta, \;  M=-J, ...,J.
\end{eqnarray}
Moreover, we have to take account of the dimension of the
representations, leading to the constraints
\be
f(J) = f(J-1) = ... = f(J-q+1) = 0
\ee
and
\be
g(-J) = g(-J+1) = ... = g(-J+q-1) = 0.
\ee
Being interested in the highest weight representations, we obtain from
(24) and (25) the following result
\begin{eqnarray}
&&f(J-(k+1)q-l)g(J-kq-l) = (k+1) \{\alpha (\frac{J-l}{q} + c)^3 +
\beta (\frac{J-l}{q} + c)^2
\nonumber \\
&&+ \gamma (\frac{J-l}{q} + c) + \delta -\frac{1}{2} [3\alpha
(\frac{J-l}{q} + c)^2 +2\beta (\frac{J-l}{q} + c) + \gamma] k \\
&&+\frac{1}{6} [3 \alpha (\frac{J-l}{q} + c) + \beta] k (2k+1) -
\frac{\alpha}{4} k^2 (k+1) \} \nonumber
\end{eqnarray}
where $l=0,1,...,q-1$ and $k=0,1,..., \frac{2J-d-l}{q}$. The
nonnegative integer $d$ introduced in the last formula has to take
specific values according to $l$ but also to $J$. These values are
summarized in the following table, $n$ denoting a nonnegative integer.

\begin{center}

 \begin{tabular}{|l||l|l|l|l|l|}
 \hline
 Table &$l=0$ &$l=1$ &$l=2$ &$...$ &$l=q-1$ \\
 \hline\hline
 $J=(qn)/2$ &$d=0$ &$d=q-1$ &$d=q-2$ &$...$ &$d=1$ \\
 $J=(qn+1)/2$ &$d=1$ &$d=0$ &$d=q-1$ &$...$ &$d=2$ \\
 $J=(qn+2)/2$ &$d=2$ &$d=1$ &$d=0$ &$...$ &$d=3$ \\
 $...$  &$...$ &$...$ &$...$ &$...$ &$...$ \\
 $J=(qn+q-1)/2$ &$d=q-1$ &$d=q-2$ &$d=q-3$ &$...$ &$d=0$ \\
 \hline
 \end{tabular}

\end{center}

Taking care of these values, we also have to constrain the real $c$ in
(21) and (27) through the conditions (26). This leads to equal to zero
the expression inside the brackets \{ . \} in (27) for
$k=\frac{2J-d-l}{q}$ or, in other words, to consider the $q$ equations
on $c$
\begin{eqnarray}
&&\alpha[c^3+\frac{3}{2q}(d-l)c^2+\frac{1}{q^2}(J^2-J(d+l)+l^2-dl+d^2)c+
\frac{1}{2q}(2J-d-l)c \nonumber \\
&&+\frac{1}{4q^3}(2J^2(d-l)-2J(d^2-l^2)+d^3-d^2l+dl^2-l^3)
+\frac{1}{4q^2}(l^2-d^2+2J(d-l))] \nonumber \\
&&+\beta[c^2+\frac{1}{q}(d-l)c+\frac{1}{3q^2}(J^2-J(d+l)+d^2-dl+l^2)
+\frac{1}{6q}(2J-d-l)] \nonumber\\
&&+\gamma (c+\frac{1}{2q}(d-l))+\delta = 0\;, \; l=0,1,...,q-1.
\end{eqnarray}
Excluding the trivial case $(\alpha, \beta, \gamma, \delta) =
(0,0,0,0)$, we notice that these equations reduce to
\be
\alpha c (c^2+J(J+1)) + \beta(c^2+\frac{1}{3}J(J+1)) + \gamma c +
\delta = 0,
\ee
for $q=1$ while for $q=2$, we are led to either
\be
\alpha = 0 \rightarrow \beta = 0, c =- \frac{\delta}{\gamma}
\ee
or
\be
\alpha \neq 0 \rightarrow \delta = \frac{\beta \gamma}{3 \alpha} -
\frac{2 \beta^3}{27 \alpha^2} , c = -\frac{\beta}{3 \alpha}
\ee
if $J = n$ and to
\be
\alpha(3c^2+\frac{1}{4}J(J+1)-\frac{1}{8})+2\beta c + \gamma = 0,
\ee
\be
\alpha(c^3+\frac{1}{4}J(J+1)c)+\beta(c^2+\frac{1}{12}J(J+1))+\gamma c
+ \delta = 0,
\ee
if $J=n+\frac{1}{2}$. Two other values of $q$ are also available
namely $q=3$ and $q=4$. We respectively obtain
\be
\alpha \neq 0, \gamma = \frac{\beta^2}{3\alpha}-\frac{1}{9}\alpha J^2
+\frac{2}{9} \alpha, \delta= \frac{\beta \gamma}{3 \alpha} -
\frac{2 \beta^3}{27 \alpha^2}, c = -\frac{\beta}{3 \alpha}
\ee
if $J=\frac{3n}{2}$,
\be
\alpha \neq 0, \gamma = \frac{\beta^2}{3\alpha}-\frac{1}{9}\alpha
J^2-\frac{2}{9}\alpha J +\frac{1}{9} \alpha, \delta= \frac{\beta
\gamma}{3\alpha} -\frac{2 \beta^3}{27 \alpha^2}, c = -\frac{\beta}{3
\alpha}
\ee
if $J=\frac{3n+1}{2}$ and
\be
\alpha \neq 0, \gamma = \frac{\beta^2}{3\alpha}-\frac{1}{9}\alpha
J^2-\frac{1}{9}\alpha J +\frac{1}{9} \alpha, \delta= \frac{\beta
\gamma}{3\alpha} -\frac{2 \beta^3}{27 \alpha^2}, c = -\frac{\beta}{3
\alpha}
\ee
if $J=\frac{3n+2}{2}$, these three contexts being associated with
$q=3$. We also have
\be
\alpha \neq 0, \gamma = \frac{\beta^2}{3\alpha}-\frac{1}{16}\alpha
J^2+\frac{3}{16} \alpha, \delta= \frac{\beta
\gamma}{3\alpha} -\frac{2 \beta^3}{27 \alpha^2}, c = -\frac{\beta}{3
\alpha}
\ee
if $J=2n$ and
\be
\alpha \neq 0, \gamma = \frac{\beta^2}{3\alpha}-\frac{1}{16}\alpha
J^2-\frac{1}{8}\alpha J +\frac{1}{8} \alpha, \delta= \frac{\beta
\gamma}{3\alpha} -\frac{2 \beta^3}{27 \alpha^2}, c = -\frac{\beta}{3
\alpha}
\ee
if $J=2n+1$, these two cases being related with $q=4$. The two other
systems related to $q=4$ i.e. those corresponding to $J =
2n+\frac{1}{2}$ and  $J =2n+\frac{3}{2}$ are incompatible ones as are
those related to $q>4$.
\par
Let us also notice that some of these representations are reducible.
For instance, if we consider the case of the usual $sl(2,R)$ algebra,
corresponding to $\alpha = 0, \beta = 0, \gamma = 2, \delta = 0$, we
obtain
\be
q=1 \rightarrow c=0
\ee
and
\be
q=2 \rightarrow c=0, J=n.
\ee
The first case (39) is associated to (see (27))
\be
f(J-k-1)g(J-k) = (k+1) \{ 2J-k \} \; , \; k=0,1,...,2J
\ee
or, in other words, to
\be
f(M-1)g(M) = (J-M+1)(J+M), M=-J,...,J.
\ee
We recognize in (42) the well known result of the angular momentum
theory [4] subtended by this $sl(2,R)$ algebra. The second case (40)
corresponds to
\be
f(J-2k-2)g(J-2k) = (k+1) \{ J-k \} \; , \; k=0,1,...,J,
\ee
\be
f(J-2k-3)g(J-2k-1) = (k+1) \{ J-k-1 \} \; , \; k=0,1,...,J-1,
\ee
or, in other words, to
\be
f(M-2)g(M) = \frac{1}{4}(J-M+2)(J+M), M=-J,-J+2,...,J,
\ee
\be
f(M-2)g(M) = \frac{1}{4}(J-M+1)(J+M-1), M=-J+1,-J+3,...,J-1.
\ee
It is then clear that the representation $(J=n,q=2)$ is in fact the
direct sum of the two (irreducible) representations
$(J=\frac{n-1}{2},q=1)$ and $(J=\frac{n}{2},q=1)$ (the eigenvalues of
the Casimir being equal).
\section{Finite-dimensional differential realizations of
$sl^{(3)}(2,R)$}
\hspace{5mm}
We now turn to the construction of the differential realizations
(expressed in terms of the real variable $x$) of the algebra
(19)-(20). In correspondence with (21)-(23), the generators of
$sl^{(3)}(2,R)$ have the following forms
\be
J_+ \equiv x^q F(D),
\ee
\be
J_0 \equiv \frac{1}{q} (D-J) + c,
\ee
\be
J_- \equiv G(D) \frac{d^q}{dx^q}
\ee
and the basis $\{ \mid J,M>, M=-J,...,J \}$ stands for the space
$P(2J)$ (cf. (8)) of monomials $\{ x^{J+M}, M=-J,...,J \}$. Moreover,
$D$ is the dilation operator
\be
D \equiv x\frac{d}{dx}.
\ee
By remembering that
\be
\frac{d^q}{dx^q} x^q = \prod_{k=1}^q (D+k) \equiv \frac{(D+q)!}{D!}
\ee
and
\be
x^q \frac{d^q}{dx^q} = \prod_{k=0}^{q-1} (D-k) \equiv
\frac{D!}{(D-q)!},
\ee
the relation (20) gives
\begin{eqnarray}
&&F(D-q)G(D-q)\frac{D!}{(D-q)!} - F(D)G(D) \frac{(D+q)!}{D!} =
\nonumber \\
&&\alpha [\frac{1}{q}(D-J)+c]^3 + \beta [\frac{1}{q}(D-J)+c]^2
\nonumber \\
&&+ \gamma [\frac{1}{q}(D-J)+c] + \delta.
\end{eqnarray}
Let us discuss this condition within the $(\alpha \neq 0)$-case first
and the $(\alpha = 0)$-case second.
a) In order to avoid singularities, we thus impose, when $\alpha \neq
0$,
\be
F(D)G(D) =
-\frac{\alpha}{4q^4}\frac{D!}{(D+q)!}(D+\lambda_1)(D+\lambda_2)
(D+\lambda_3)(D+\lambda_4)
\ee
and obtain the following system on the real unknows $\lambda_1,...,
\lambda_4$
\be
\lambda_1 + \lambda_2 + \lambda_3 + \lambda_4 = 4qc + 2q - 4J +
\frac{4}{3} \frac{\beta}{\alpha} q,
\ee
\begin{eqnarray}
&&\lambda_1\lambda_2 + \lambda_1\lambda_3 + \lambda_1\lambda_4 +
\lambda_2\lambda_3 + \lambda_2\lambda_4 + \lambda_3\lambda_4 = 6q^2c^2
+ 6q^2c - 12 qJc \nonumber \\
&&+4\frac{\beta}{\alpha}q^2c + 6J^2 - 6qJ - 4\frac{\beta}{\alpha}qJ +
q^2 + 2 \frac{\beta}{\alpha}q^2 + 2\frac{\gamma}{\alpha}q^2,
\end{eqnarray}
\begin{eqnarray}
&&\lambda_1\lambda_2\lambda_3 + \lambda_1\lambda_2\lambda_4 +
\lambda_1\lambda_3\lambda_4 + \lambda_2\lambda_3\lambda_4 = \nonumber
\\
&&4q^3c^3 - 12 q^2Jc^2 + 6q^3c^2 + 4\frac{\beta}{\alpha}q^3c^2 +
12qJ^2c - 12q^2Jc \nonumber \\
&&+2q^3c - 8\frac{\beta}{\alpha}q^2Jc + 4\frac{\beta}{\alpha}q^3c +
4\frac{\gamma}{\alpha}q^3c - 4J^3 + 6qJ^2 - 2q^2J \nonumber \\
&&+4\frac{\beta}{\alpha}qJ^2-4\frac{\beta}{\alpha}q^2J + \frac{2}{3}
\frac{\beta}{\alpha}q^3 - 4\frac{\gamma}{\alpha}q^2J +
2\frac{\gamma}{\alpha}q^3 + 4\frac{\delta}{\alpha}q^3.
\end{eqnarray}
Let us recall that we are interested in finite-dimensional (=$2J+1$)
realizations only. This means that the cases $q=1$ and $q=2$ are the
only possibilities in accordance with
\be
q=1 \rightarrow \lambda_1 = 1, \lambda_2 = -2J
\ee
and
\be
q=2 \rightarrow \lambda_1 = 1, \lambda_2 = 2, \lambda_3 = -2J,
\lambda_4 = -2J+1.
\ee
In the first context ($q=1$), the equations (55) and (56) fix
$\lambda_3$ and $\lambda_4$ as follows
\be
\lambda_3 = 2c+\frac{1}{2}-J+\frac{2}{3}\frac{\beta}{\alpha} +
\frac{\epsilon}{2}, \lambda_4 =
2c+\frac{1}{2}-J+\frac{2}{3}\frac{\beta}{\alpha} -
\frac{\epsilon}{2}
\ee
with
\be
\epsilon^2 = 1 + \frac{16}{9} \frac{\beta^2}{\alpha^2} - 4J(J+1) -
8c^2 - \frac{16}{3} \frac{\beta}{\alpha} c - 8 \frac{\gamma}{\alpha}
\ee
while the equation (57) coincides with (29). In the second context
($q=2$), we are led to (31) supplemented by
\be
\gamma = \frac{\beta^2}{3\alpha} - \frac{\alpha}{4} J^2 -
\frac{\alpha}{4} J + \frac{\alpha}{8} .
\ee
\par
b) The case $\alpha = 0$ is more simple and has already been analyzed
[5]. For self-consistency, we recall the main results i.e.
\be
F(D)G(D) = -\frac{\beta}{3q^3} \frac{D!}{(D+q)!} (D+\lambda_1)
 (D+\lambda_2) (D+\lambda_3)
\ee
where the only possible finite-dimensional (=$2J+1$) realization is
associated with $q=1$ and corresponds to
\be
\lambda_1 = 1, \lambda_2 = -2J, \lambda_3 =
-J+3c+\frac{3\gamma}{2\beta}+\frac{1}{2},
\ee
the real $c$ being fixed through
\be
\beta c^2 + \gamma c + \delta + \frac{\beta}{3} J(J+1) = 0.
\ee
\par
Now that both cases have been considered, let us conclude this Section
by noticing that some realizations (namely the ones corresponding to
$q=2$ without the condition (62) and the ones associated with $q=3,4$)
are missing with respect to the representations developed in the
previous Section. Indeed, the relation (53) we have imposed is more
constraining because it is a relation between operators independently
of the basis on which they are supposed to act. If we take account of
this basis i.e. in this case $P(2J) = \{ x^{J+M}, M=-J,...,J \}$, we
can recover all the cases previously discussed. For example, in the
context $q=3$,
$J=\frac{3}{2}$, we can consider
\be
J_+ = -\frac{1}{6} f(-\frac{3}{2})x^3(D-1)(D-2)(D-3),
\ee
\be
J_0 = \frac{1}{3}D - \frac{1}{2} - \frac{\beta}{3\alpha},
\ee
\be
J_- = \frac{1}{6} g(\frac{3}{2}) \frac{d^3}{dx^3}
\ee
with
\be
f(-\frac{3}{2})g(\frac{3}{2}) = \frac{\alpha}{9}.
\ee
It is then easy to convince ourselves that these operators generate
$sl^{(3)}(2,R)$ with $\gamma =
\frac{\beta^2}{3\alpha}-\frac{\alpha}{36}$ and $\delta =
\frac{\beta^3}{27\alpha^2}-\frac{\beta}{108}$ but on the space $P(3)$
only (the relation (53) being trivially not satisfied except on this
space). However it has to be stressed that the relations corresponding
to (66)-(68) but in the general context become really heavy when the
value of $J$ increases.
\section{Two examples}
\hspace{5mm}
We first consider the prototype of QES systems i.e. the so-called
sextic oscillator [1,2] and then turn to a more physical example: the
SHG problem.
\subsection{The sextic oscillator}
\hspace{5mm}
This system is characterized by the following potential
\be
V(y) = a^2 y^6 + 2ab y^4 + (b^2-2ap-8aj-3a) y^2
\ee
with $a(\neq 0), b \in R$ and $p=0,1$ while $j$ is the quantum number
appearing in (4). With
\be
x = y^2
\ee
and
\be
\chi = - \int{(\frac{a}{2}x+\frac{b}{2}-\frac{p}{2x})dx},
\ee
we can be convinced of its QES character via the form (7) (up to a
translation)
\be
H = J_+ + J_0 + J_- + 2bp + 4bj + b
\ee
and
\be
c^{0-}=-4, c^+=-4a, c^0=4b, c^-=-(4j+2+4p),
\ee
the other $c's$ being equal to zero. Without loss of generality, we
can put $b=\frac{1}{4}$ (in order to recover (19)) and obtain, through
(11)-(13), the relation (20) with
\be
\alpha = 0, \beta = 48a, \gamma = 32a(p+j), \delta = -16aj(j+1).
\ee
The case $q=1$ is thus the only one to be available. We actually have
\be
J_0 \mid J,M> = (M+c) \mid J,M>,
\ee
\be
J_+ \mid J,M> = f(M) \mid J,M+1>,
\ee
\be
J_- \mid J,M> = g(M) \mid J,M-1>,
\ee
with
\be
f(M-1)g(M) = (J-M+1)(J+M)(48ac+16aM+16ap+16aj-8a).
\ee
Moreover, the parameter $c$ is fixed according to
\be
c = -\frac{1}{3}(p+j) \pm \frac{1}{3} \sqrt{(p+j)^2-3J(J+1)+3j(j+1)}
\ee
leading to constrain $J$ through
\be
J \leq -\frac{1}{2} + \frac{1}{6}\sqrt{36j(j+1)+12(p+j)^2+9}
\ee
in order to ensure the reality of $c$. Because the space $P(2J)$ is
preserved, we just have to equal to zero the determinant of the
following matrix
\begin{eqnarray}
&&\rm M = \sum_{k=1}^{2J+1} (E+J-k-c-\frac{p}{2}-j+\frac{3}{4})e_{k,k}
\nonumber \\
&&-\sum_{k=1}^{2J}g(-J+k) e_{k,k+1} -
\sum_{k=1}^{2J}f(-J+k-1)e_{k+1,k}
\end{eqnarray}
in order to find the energies. In the matrix (82), the notation
$e_{k,l}$ stands for a $(2J+1)$-dimensional matrix where 1 is at the
intersection of the $k^{th}$ row and the $l^{th}$ column, the other
elements being 0. For example, when $j=\frac{1}{2}$, we have
\be
J = 0, \frac{1}{2},
\ee
according to (81) while the relation (80) gives
\be
c = -\frac{p}{3} - \frac{1}{6} \pm
\frac{1}{3}\sqrt{(p+\frac{1}{2})^2+\frac{9}{4}}
\ee
if $J=0$ and
\be
c=0 \; \; or \; \; c = -\frac{1}{3} (2p+1)
\ee
if $J = \frac{1}{2}$. In the case (84), the energies are
\be
E=c+\frac{p}{2}+\frac{3}{4} \rightarrow
E=0.0428932;0.0562872;1.1103796;1.4571067
\ee
and in the case (85), the resolution of the vanishing determinant
associated with (82) leads to
\be
E=\frac{3}{4} \pm \frac{1}{2}\sqrt{1+32a}; E=\frac{5}{4} \pm
\frac{1}{2}\sqrt{1+96a}
\ee
if $c=0$ and
\be
E=\frac{5}{12} \pm \frac{1}{2}\sqrt{1-32a}; E=\frac{1}{4} \pm
\frac{1}{2}\sqrt{1-96a}
\ee
if $c=-\frac{1}{3}$; $c=-1$. Only the values given in (87) correspond
to the previously obtained ones [2]. This is indeed a general result
that the known energies [2] are recovered through our approach when
$c=0, J=j$. The other contexts ($J<j$) or ($J=j, c=-\frac{2}{3}(p+j)$)
lead to supplementary new values of the energy.
\par
Let us analyze more deeply this result by going to the differential
realization (47)-(49) i.e.
\be
J_+ = x F(D),
\ee
\be
J_0 = D-J+c,
\ee
\be
J_- = G(D) \frac{d}{dx},
\ee
where (cf. (63) and (64))
\be
F(D)G(D) = -16a (D-2J)(D-J+3c+\frac{1}{2}+p+j).
\ee
In order to preserve the space $P(2J)$, let us make the choice
(without loss of generality, this freedom being due to the fact that
$sl^{(3)}(2,R)$ is defined up to an automorphism [5])
\be
G(D) = -4(D-J+3c+\frac{1}{2}+p+j).
\ee
In that case, the Hamiltonian (73) is realized as
\begin{eqnarray}
&&H =-4 x \frac{d^2}{dx^2} +
[4ax^2+x-4(-J+3c+\frac{1}{2}+p+j)]\frac{d}{dx} \nonumber\\
&&-8aJx+\frac{p}{2}+\frac{1}{4}+j-J+c.
\end{eqnarray}
This form is analog to (2) and we propose to write it in the Schr\"
odinger form (1) through the changes (3) i.e.
\be
x = y^2,
\ee
\be
\psi =
exp(\int{(\frac{1}{8}-2ax+\frac{1}{2}(J-3c-p-j)\frac{1}{x})dx})\phi.
\ee
The potential obtained in this manner is given by
\begin{eqnarray}
&&V(y)= a^2 y^6 +\frac{1}{2}a y^4 +
(\frac{1}{16}-6aJ-2aj-6ac-2ap-3a)y^2 \nonumber \\
&&+\frac{1}{2}(j-J-c) + (J-3c-p-j)(J-3c-p-j+1)\frac{1}{y^2}.
\end{eqnarray}
Compared with (70), this expression actually reduces to the sextic
oscillator potential iff $c=0$ and $J=j$. For other values of the
parameters (i.e. the already cited $(J<j)$ and $(J=j, c
=-\frac{2}{3}(p+j))$, the new eigenvalues of the problem (appearing
for $j=\frac{1}{2}$ in (86) and (88)) do correspond to another model,
namely the {\it radial} sextic oscillator as shown by (97).

\subsection{The second harmonic generation}
\hspace{5mm}
This nonlinear optical process as well as others such as coherent
spontaneous emission and down conversion [6] can be described by the
following effective Hamiltonian
\be
H = a_1^{\dagger}a_1 + 2 a_2^{\dagger}a_2 +g(a_2^{\dagger}a_1^2 +
(a_1^{\dagger})^2a_2)
\ee
with cubic terms in the (independent) bosonic creation and
annihilation operators. The Hamiltonian (98) has already been
recognized [7] as a QES model. We confirm such a result by using the
technique developed in Section 2. Indeed following Section 2, we
propose to define
\be
J_+ = a_2^{\dagger}a_1^2,
\ee
\be
J_0 = \frac{1}{3} (a_2^{\dagger}a_2-a_1^{\dagger}a_1),
\ee
\be
J_- = (a_1^{\dagger})^2a_2
\ee
such that the algebra (19)-(20) is satisfied with
\be
\alpha = 0, \beta = -12, \gamma =0, \delta = \frac{1}{3}N^2+N.
\ee
In the last expression, $N$ is the invariant
\be
N = a_1^{\dagger}a_1 + 2 a_2^{\dagger}a_2
\ee
satisfying
\be
[N , J_0] = [N , J_{\pm}] = 0.
\ee
Once again, the values (102) are typical of the $q=1$-representation
only so that the relations (76)-(78) are the ones to be taken care of
but with
\be
f(M-1)g(M) = (J-M+1)(J+M)(-12c-4M+2).
\ee
The parameter $c$ is fixed according to
\be
c^2 =-\frac{1}{3}J(J+1) + \frac{1}{36}N^2+\frac{1}{12}N
\ee
and its reality is ensured if
\be
J \leq -\frac{1}{2} + \frac{1}{2}\sqrt{\frac{1}{3}N^2+N+1}.
\ee
It is then possible to determine the energies by putting to zero the
determinant of a $(2J+1)$ by $(2J+1)$ matrix analog to (82) as well as
it is possible to determine them through the differential realization
(89)-(91). In this case, we have
\be
F(D)G(D) = 4 (D-2J)(D-J+3c+\frac{1}{2})
\ee
and choosing
\be
G(D) = 1
\ee
the Hamiltonian (98) becomes
\begin{eqnarray}
&&H = N + 4gx^3\frac{d^2}{dx^2} +
g(1+12(-J+c+\frac{1}{2})x^2)\frac{d}{dx} \nonumber \\
&&+4g(-J+2J^2-6Jc)x.
\end{eqnarray}
With the respective changes of variables and wavefunctions
\be
x = -\frac{1}{gy^2}, \psi =
exp(-\int{(\frac{1}{8x^3}+\frac{3}{2}(c-J)\frac{1}{x})dx}) \phi
\ee
we can put (110) on the Schr\" odingerlike form (1) with
\be
V(y) =
\frac{g^4}{16}y^6+\frac{3}{2}g^2(c-J-\frac{1}{2})y^2
+(J+3c)(J+3c+1)\frac{1}{y^2}+N.
\ee
Once again this is typical of a radial sextic oscillator and the QES
characteristics of the second harmonic generation is thus
proved.  The determination of the energies is then possible without
any difficulty [2]. For example, when $N=4$, we have
\be
J=0 \rightarrow E = 4,
\ee
\be
J=\frac{1}{2} \rightarrow E = 4 \pm g\sqrt{2\sqrt{19}},
\ee
\be
J=1 \rightarrow E = 4, 4 \pm 4g.
\ee
Only the values (115) correspond to known ones, the values (113),
(114) coming from other models. This is a general result in the sense
that the energies of SHG are the ones of the Schr\" odinger potential
(112) with
\be
c = J - \frac{N}{3}
\ee
and
\be
J = \frac{N}{4} \; \; or \; \; J = \frac{N-1}{4}
\ee
according to even or odd values of $N$. The SHG potential thus writes
\be
V(y) = \frac{g^4}{16} y^6 - \frac{g^2}{4} (2N+3)y^2 + N
\ee
and exactly coincides with the potential (42) of Ref. [7]. In terms of
the operators (4) (with $j=J$), this gives
\be
H = gj_+ (j_0 + \frac{1}{2} - \frac{N}{4}) - 4gj_- + N
\ee
if $N$ is even and
\be
H = gj_+ (j_0 - \frac{1}{4} - \frac{N}{4}) - 4gj_- + N
\ee
if $N$ is odd.

\section{Conclusions}
\hspace{5mm}
We have developed a general method based on the polynomial
deformations of the Lie algebra $sl(2,R)$ in order to exhibit the QES
characteristics of a Hamiltonian. We have applied this method to two
examples: one more theoretical -the sextic oscillator- and one more
physical -the second harmonic generation-. In both cases, a finite
number of energies as well as eigenfunctions are determined through
the finite-dimensional representations -or, in an equivalent way,
through the realizations- of these polynomial deformations. Some of
these energies (and eigenfunctions) were previously known, not the
others. It seems, through the analysis of the two examples, that these
previously unknown energies do not correspond to the same model but to
another one being closely related to the first one. In some cases, the
comparison between these new and old models could be interesting, with
respect to experimental data in particular.
\par
The main advantage of the method we have proposed is that it can be
systematically applied to any Hamiltonian written in terms of a
raising and a lowering operator. Numerous physical Hamiltonians are of
that type. In particular, we plan to analyze one of them: the
so-called Lipkin-Meshkov-Glick Hamiltonian [8] of specific interest in
nuclear physics. Being based on a $(\alpha \neq 0)$ polynomial
deformation, its analysis is more delicate but also richer in the
number of available representations [9].
\par
The main drawback is that it is limited to $sl(2,R)$ and its
deformations when we know that some QES models need more extended
algebras. However, the method we have presented here can also be
generalized to these extended algebras as well as superalgebras. We
also plan to come back on these points in the near future.

\vspace{5mm}

{\bf ACKNOWLEDGMENTS}
\par
I would like to warmly thank Prof. J. Beckers (University of Li\`ege,
Belgium) for fruitful discussions. Thanks are also due to Prof. A.
Klimov (University of Guadalajara, Mexico) for useful information on
the SHG model and for pointing out the Ref. [7] in particular.

\end{document}